\documentclass[sigconf]{acmart}

\usepackage{tikz}
\usetikzlibrary{positioning}
\usepackage{fontawesome}
\usepackage{comment}
\usepackage[many]{tcolorbox}

\AtBeginDocument{%
  \providecommand\BibTeX{{%
    \normalfont B\kern-0.5em{\scshape i\kern-0.25em b}\kern-0.8em\TeX}}}

\copyrightyear{2021}
\acmYear{2021}
\setcopyright{acmlicensed}
\acmConference[CSCW '21 Companion]{Companion Publication of the 2021 Conference on Computer Supported Cooperative Work and Social Computing}{October 23--27, 2021}{Virtual Event, USA}
\acmBooktitle{Companion Publication of the 2021 Conference on Computer Supported Cooperative Work and Social Computing (CSCW '21 Companion), October 23--27, 2021, Virtual Event, USA}
\acmPrice{15.00}
\acmDOI{10.1145/3462204.3481742}
\acmISBN{978-1-4503-8479-7/21/10}

\begin{document}\sloppy

\title[Appropriate Fairness Perceptions]{Appropriate Fairness Perceptions? On the Effectiveness of Explanations in Enabling People to Assess the Fairness of Automated Decision Systems}

\author{Jakob Schoeffer}
\orcid{0000-0003-3705-7126}
\email{jakob.schoeffer@kit.edu}
\affiliation{%
  \institution{Karlsruhe Institute of Technology (KIT)}
  \country{Germany}
}

\author{Niklas Kuehl}
\orcid{0000-0001-6750-0876}
\email{niklas.kuehl@kit.edu}
\affiliation{%
  \institution{Karlsruhe Institute of Technology (KIT)}
  \country{Germany}
}


\begin{abstract}
    It is often argued that one goal of explaining automated decision systems (ADS) is to facilitate positive perceptions (e.g., fairness or trustworthiness) of users towards such systems.
    This viewpoint, however, makes the implicit assumption that a given ADS is fair and trustworthy, to begin with.
    If the ADS issues unfair outcomes, then one might expect that explanations regarding the system's workings will reveal its shortcomings and, hence, lead to a \emph{decrease} in fairness perceptions.
    Consequently, we suggest that it is more meaningful to evaluate explanations against their effectiveness in enabling people to appropriately assess the quality (e.g., fairness) of an associated ADS.
    We argue that for an effective explanation, perceptions of fairness should increase \emph{if and only if} the underlying ADS is fair.
    In this in-progress work, we introduce the desideratum of \emph{appropriate fairness perceptions}, propose a novel study design for evaluating it, and outline next steps towards a comprehensive experiment.
  %
  %
  %
  %
  %
  %
\end{abstract}



\maketitle

\section{Introduction}\label{sec:introduction}

Automated decision systems (ADS) have become ubiquitous in consequential decision-making, spanning domains such as lending, hiring, and policing \cite{townson2020ai,kuncel2014hiring,heaven2020predictive}.
These systems are typically based on sophisticated machine learning (ML) models that are difficult to interpret and assess (\emph{black-box models}); particularly for people \emph{affected} by automated decisions.
This is problematic, both legally and morally.
The research community---partially as a response to novel laws and regulations---has suggested explanations as a means to remedy the negative consequences of opaque ADS.

A natural question that arises is how to evaluate the effectiveness of such explanations.
The answer will depend, among others, on the nature of stakeholders, i.e., the recipients of explanations.
%
One can distinguish two relevant, yet fundamentally different, stakeholders: decision-makers (who use ADS for decision support) and decision-subjects (e.g., loan applicants).
Typically, the former are primarily interested in making ``good'' decisions w.r.t. some predefined target.
If ground-truth labels are available for a given task, this might, e.g., translate to maximizing classification accuracy.
In that case, explanations' contribution towards making more accurate decisions could be interpreted as their effectiveness.
Decision-subjects, on the other hand, are \emph{confronted} with decisions about themselves (e.g, automated loan denial) that they seek to comprehend.
%
%
Typical questions they might ask are, \emph{Why did the ADS make this decision?} or \emph{Am I being treated fairly by the ADS?}
%
In this work, contrary to most prior research, we take the perspective of (potential) decision-subjects and examine what effective explanations
of ADS
might entail.
Specifically, we aim to understand which explanation styles can (or cannot) enable decision-subjects to assess the fairness of ADS.
Note that while vetting of ADS should in most cases \emph{not} be the responsibility of decision-subjects, we argue that they may still greatly benefit from appropriate fairness perceptions in case of deficient (or selective) auditing by technical staff or regulatory agencies.

\paragraph{Related Work}
Prior work has already formulated various desiderata for explanations of ADS \cite{adadi2018peeking,lipton2018mythos,doshi2017towards,langer2021we}.
Very recently, \citet{wang2021explanations}, after thoroughly reviewing and summarizing existing research, put forward that explanations should a) improve people's understanding of an ADS; b) help people recognize the uncertainty underlying a prediction and nudge people to rely on the ADS more on high confidence predictions when the ADS’s confidence is calibrated; c) empower people to trust the ADS appropriately.
However, these desiderata are formulated from the perspective of decision-makers, i.e., \emph{not} the decision-subjects.
This might explain why a fairness dimension is missing, because stakes for decision-subjects are presumably much higher w.r.t. (un)fairness of ADS than for decision-makers.

Another, relatively new, line of research has been looking at how people's perceptions of fairness and trustworthiness towards ADS change based on the provision of different explanations \cite{binns2018s,dodge2019explaining,lee2018understanding,lee2017algorithmic,wang2020factors,lee2019procedural,kizilcec2016much}.
These works, however, either a) examine perceptions of decision-makers (e.g., \cite{wang2020factors}), or b) consider ADS that are externally given
(e.g., \cite{kizilcec2016much,lee2017algorithmic,lee2018understanding,lee2019procedural}) or hypothetical altogether (e.g., \cite{binns2018s}).
%
A caveat about examining externally-given ADS involves that we cannot intervene w.r.t. the quality (e.g., fairness) of the system.
Hence, important issues on whether more positive perceptions (e.g., of trustworthiness or fairness) due to explanations will happen \emph{independently} of the quality of the ADS---which would be highly undesirable and a potential sign for explanations that are too \emph{persuasive} \cite{chromik2020mind}---remain unanswered.

Perhaps the most related to our work is a study by \citet{dodge2019explaining}, which is, to the best of our knowledge, the only study to a) employ an ADS based on real-world data, b) randomize the ``fairness'' of the ADS, and c) examine perceptions in a scenario-based experiment.
However, we have identified several aspects that demand follow-up work like ours: First, the authors employ only two specific notions of fairness based on pre-processing the ADS's training data \cite{calmon2017optimized} as well as \emph{case-specific disparate impact} \cite{grgic2018human,calders2013unbiased}; second, the use case of recidivism prediction might not be as relatable by laypeople; third, study participants' fairness judgment is assessed based on a single-item construct only.

\paragraph{Our Contributions}
Our contribution with this (in-progress) work is threefold: First, we assess the effectiveness of ADS explanations from the perspective of decision-subjects (as opposed to decision-makers), which is still under-researched.
To that end, second, we formulate a new desideratum for explanations, namely \emph{appropriate fairness perceptions}---which is satisfied if people's fairness perceptions towards ADS are high if the system is indeed fair; and they are low if the system behaves unfairly.
Third, we introduce a study design to evaluate our criterion for different explanation styles.
Specifically, we propose a between-subject experiment where 50\% of participants each are shown output from a fair and an unfair ADS, respectively.
Through randomization, this allows to compare fairness perceptions in both groups and assess whether they are appropriate for any given explanation style.
As compared to the work by \citet{dodge2019explaining}, we use different fairness notion(s), a different use case (lending), and employ a more nuanced way of assessing perceptions through several multi-item constructs, mostly drawn and adapted from the organizational justice literature \cite{colquittorganizational}.

\section{Background}\label{sec:background}



Before we detail our proposed study setup, we introduce the employed explanation styles and provide some background on a fundamental question that our work is based upon: \emph{What constitutes a fair ADS?}

\subsection{Explanation Styles}\label{sec:explanation_styles}

Much research has been trying to categorize different explanation styles (e.g., \cite{wachter2017right,adadi2018peeking,edwards2017slave, guidotti2018survey, binns2018s}).
%
%
In this work, we employ the condensed model-agnostic taxonomy introduced by \citet{binns2018s} because of its manageable dimensions and supposed GDPR-compliance \cite{binns2018s}.
Note that our study design is readily applicable to other explanation styles as well.
The four different explanation styles proposed by \citet{binns2018s} are given as follows; where the concrete explanations can, e.g., be taken from our previous work \cite[p. 10--12]{schoeffer2021study} and---adjusted for the use case---from \citet[p. 6]{binns2018s}:

\begin{itemize}
    \item \emph{Input influence.\ } A list of input variables (i.e., features) used by the ADS, alongside a measure of how (e.g., positively or negatively) they influence the decision.
    \item \emph{Sensitivity.\ } For each input variable used in the decision, how much would its value have to differ in order to change the ADS's decision.
    \item \emph{Case-based.\ } A case of the underlying ML model's training data which is most similar to the decision being explained.
    \item \emph{Demographic.\ } Aggregate statistics on the outcome distribution for people in the same demographic categories as a given decision-subject (e.g., age, gender, income).
\end{itemize}


\subsection{Determining Fairness of ADS}\label{sec:determining_fairness_of_ADS}

A fundamental question we need to ask with respect to appropriate fairness perceptions is, \emph{In which cases do we want an ADS to be perceived as (un)fair?} or, similarly, \emph{What constitutes a fair ADS?}
First and foremost, we need to acknowledge that there is not \emph{one} way of deciding whether a given ADS (or rather, the underlying ML model) is fair or not.
In fact, there exist many different criteria in the algorithmic fairness literature \cite{mehrabi2021survey}, some of which are contradicting each other \cite{chouldechova2017fair}.
Examples of commonly-regarded fairness notions are, e.g., \emph{demographic parity} \cite{mehrabi2021survey}, \emph{equality of opportunity} \cite{hardt2016equality}, \emph{meritocratic fairness} \cite{kearns2017meritocratic,schoeffer2021ranking}, or \emph{fairness through awareness} \cite{dwork2012fairness}.
From the (statistical) definitions of these notions, it seems likely that a) some notions are more relevant in specific contexts than others (e.g., lending vs. hiring), and b) different types of information/explanations will be needed by decision-subjects for spotting violations, depending on the adopted (un)fairness notion.
From a technical perspective, our proposed study design can handle any (un)fairness notion provided that we can construct an unfair ADS as well as a fair baseline---we cover this in more detail in Sec.~\ref{sec:construction_of_fair_and_unfair_ADS}.
%
%
%

Acknowledging the fact that \emph{fairness} is a contested concept \cite{mulligan2019thing}, researchers have also turned to evaluating people's perceptions towards ADS as a means to determine these systems' fairness.
Particularly the organizational justice literature (e.g., \cite{colquittorganizational}) appears to offer a rich set of suitable constructs, which have been increasingly adopted by HCI and CSCW scholars recently \cite{binns2018s,lee2019procedural,watkinsa2020tension}.
%
%
We, too, employ these constructs to measure fairness perceptions.
More details will be given in Sec.~\ref{sec:comparing_fairness_perceptions}.










\section{Proposed Study Setup}\label{sec:proposed_study_setup}

\begin{figure*}[t]
\centering
\resizebox{0.6\textwidth}{!}{
\begin{tikzpicture}  
  [scale=1] 
   
\node (n0) at (0,0) {\Huge \faUsers}; 

\node[style={rectangle, draw=black, fill=gray!10, minimum width=3cm, minimum height=0.5cm, text centered, anchor=east}] (n1) at (-0.5,-1) {Fair ADS}; 

\node[style={rectangle, draw=black, fill=gray!40, minimum width=3cm, minimum height=0.5cm, text centered, anchor=west}] (n2) at (0.5,-1) {Unfair ADS}; 


\node[style={rectangle, draw=black, fill=gray!10, minimum width=3cm, minimum height=0.5cm, text centered, anchor=west}] (n3) at (-5,-2) {Input Influence};

\node[style={rectangle, draw=black, fill=gray!10, minimum width=3cm, minimum height=0.5cm, text centered, anchor=west}] (n4) at (-5,-2.75) {Sensitivity}; 

\node[style={rectangle, draw=black, fill=gray!10, minimum width=3cm, minimum height=0.5cm, text centered, anchor=west}] (n5) at (-5,-3.5) {Case-Based};

\node[style={rectangle, draw=black, fill=gray!10, minimum width=3cm, minimum height=0.5cm, text centered, anchor=west}] (n6) at (-5,-4.25) {Demographic}; 


\node[style={rectangle, draw=black, fill=gray!40, minimum width=3cm, minimum height=0.5cm, text centered, anchor=east}] (n7) at (5,-2) {Input Influence};

\node[style={rectangle, draw=black, fill=gray!40, minimum width=3cm, minimum height=0.5cm, text centered, anchor=east}] (n8) at (5,-2.75) {Sensitivity}; 

\node[style={rectangle, draw=black, fill=gray!40, minimum width=3cm, minimum height=0.5cm, text centered, anchor=east}] (n9) at (5,-3.5) {Case-Based};

\node[style={rectangle, draw=black, fill=gray!40, minimum width=3cm, minimum height=0.5cm, text centered, anchor=east}] (n10) at (5,-4.25) {Demographic}; 


\draw[->, -latex] (n0) -| (n1);
\draw[->, -latex] (n0) -| (n2);

\draw[->, -latex] (-6,-1) |- (n3.west);
\draw[->, -latex] (-6.25,-1) |- (n4.west);
\draw[->, -latex] (-6.5,-1) |- (n5.west);
\draw[->, -latex] (n1.west) -- (-6.75,-1) |- (n6.west);

\draw[->, -latex] (6,-1) |- (n7.east);
\draw[->, -latex] (6.25,-1) |- (n8.east);
\draw[->, -latex] (6.5,-1) |- (n9.east);
\draw[->, -latex] (n2.east) -- (6.75,-1) |- (n10.east);

\end{tikzpicture}
}
\caption{Graphical framework of our proposed study setup.}
\label{fig:study_setup}
\end{figure*}
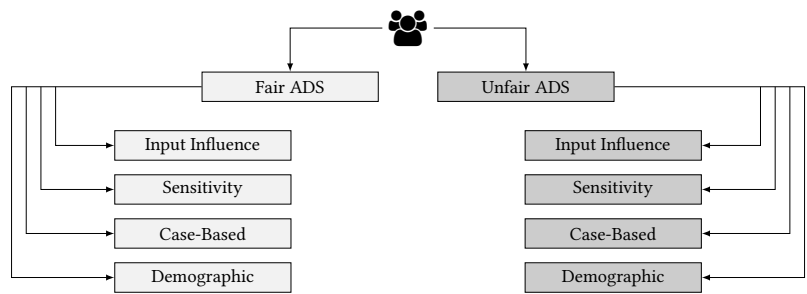

To examine which explanation styles are effective in facilitating appropriate fairness perceptions, we propose a scenario-based online study with randomly selected adult participants.
%
As mentioned in Sec.~\ref{sec:determining_fairness_of_ADS}, it will be important to examine and compare different contexts and communities.
For clarity of exposition, however, we introduce our proposed study setup only for the context of \emph{lending}.
%
Specifically, we aim to confront study participants with several scenarios where a person was denied a loan; the proposed introductory text for each scenario is given in Fig.~\ref{fig:introductory_text}.
\begin{figure}[t]
    \centering
    \resizebox{0.8\columnwidth}{!}{
    \begin{tcolorbox}[enhanced jigsaw, sharp corners, colframe=black, boxrule=0.5pt]
    A finance company offers loans on real estate in urban, semi-urban, and rural areas. A potential customer first applies online for a specific loan, and afterward, the company assesses the customer's eligibility for that loan.

    An individual applied online for a loan at this company. The company denied the loan application. The decision to deny the loan was made by an automated decision system.
    \end{tcolorbox}
    }
    \caption{Proposed introductory text (before explanations).}
    \label{fig:introductory_text}
\end{figure}
We propose a $2\times4$ between-subject design with the following conditions: First, we randomly assign study participants the fair or the unfair ADS; then, we provide them with one of four explanation styles (see Fig.~\ref{fig:study_setup}).
For simplicity, we might also consider conducting four individual sub-experiments, one for each explanation style.
Eventually, we measure the participants' perceptions of fairness given their condition, and we compare the results across conditions.
Importantly, note that for any adopted (un)fairness notion, we need to ensure that study participants actually subscribe to it; i.e., if study participants (hypothetically) knew that the ADS violates the given notion, they would find this troubling.
We will test this within our questionnaires.

\subsection{Construction of Fair and Unfair ADS}\label{sec:construction_of_fair_and_unfair_ADS}

%

For the context of lending, we base our analyses on a publicly available dataset on home loan application decisions.\footnote{\url{https://www.kaggle.com/altruistdelhite04/loan-prediction-problem-dataset} (last accessed August 4, 2021)}
%
%
The cleaned dataset consists of 480 labeled (loan $Y/N$) observations and includes the following features: \textit{applicant income, co-applicant income, credit history, dependents, education, gender, loan amount, loan amount term, marital status, property area, self-employment}.
We use this dataset for training our fair and unfair classifiers/ADS.
%
%

As mentioned in Sec.~\ref{sec:determining_fairness_of_ADS}, our study setup can (and should) handle different (un)fairness notions.
Due to its prominence in the literature as well as straightforward operationalization, we will first employ the notion of \emph{demographic (dis)parity}, i.e., the idea that the likelihood of loan approval should be the same regardless of whether a person is member of a protected group (e.g., females) or not \cite{mehrabi2021survey}.
This parity constraint can be readily enforced when training a classifier, whose output will then serve as the \emph{fair ADS} baseline condition.
Note that we could also (continuously) relax this parity constraint to satisfy, e.g., the so-called ``80\%-rule'' \cite{biddle1995disparate}.
As for the \emph{unfair ADS} condition, we can modify the classifier such that the decision solely (or at least mostly) depends on an applicant's gender, implying full disparity (see, e.g., \cite{schoeffer2021ranking} for an implementation).
After training the fair and unfair classifiers, we generate automated decisions (loan $Y/N$) for all observations from a predefined holdout set each.
We then randomly sample applicants from the subset of rejected loans to use them as scenarios in our questionnaires.

Apart from demographic parity, we plan on evaluating explanations based on other fairness notions as well, e.g., \emph{fairness through awareness} (``treating similar individuals similarly'') \cite{dwork2012fairness} or \emph{meritocratic fairness} \cite{schoeffer2021ranking,kearns2017meritocratic}.
However, for these notions, constructing a fair baseline ADS is somewhat less obvious and requires additional groundwork, such as supplementary upstream experiments.

\subsection{Comparing Fairness Perceptions}\label{sec:comparing_fairness_perceptions}

After constructing the fair and unfair ADS, we can separately generate the explanations from Sec.~\ref{sec:explanation_styles} for both models and provide them to study participants in the respective conditions---see \cite[p. 10--12]{schoeffer2021study} or \cite[p. 6]{binns2018s} for concrete examples.
For \emph{input influence}, we compute permutation feature importance \cite{breiman2001random}.
For \emph{sensitivity}, we compute counterfactuals \cite{karimi2021algorithmic} per feature.
For \emph{case-based} explanations, we provide the most similar observation from the training data, based on the Euclidean distance in the (scaled) feature space.
Lastly, \emph{demographic} explanations are generated just as introduced in Sec.~\ref{sec:explanation_styles}.
Note that explanations will differ (except \emph{case-based}) for the fair and the unfair ADS, as desired.
%



To measure study participants' fairness perceptions towards the ADS, we use constructs from the organizational justice literature \cite{colquittorganizational}, as mentioned in Sec.~\ref{sec:determining_fairness_of_ADS}.
Specifically, Colquitt and Rodell \cite{colquittorganizational,colquitt2015measuring} propose four dimensions that make up ``overall fairness perceptions.''
The associated (multi-item) constructs, including some intuition, are:
\begin{itemize}
    \item \emph{Distributive justice.\ } Fairness perceptions regarding decision outcomes (e.g., \emph{Are the outcomes justified, given the applicants' qualifications?})
    \item \emph{Procedural justice.\ } Fairness perceptions regarding the decision-making procedures (e.g., \emph{Are the decision-making procedures applied consistently?})
    \item \emph{Interpersonal justice.\ } Fairness perceptions regarding interactions/communication with the ADS (e.g., \emph{Are the applicants treated with respect?})
    \item \emph{Informational justice.\ } Fairness perceptions regarding justification and truthfulness of explanations regarding decision-making procedures (e.g., \emph{Does the ADS explain decision-making procedures thoroughly?})
\end{itemize}
%
%
Finally, we need to compare the results between the fair and the unfair ADS.
%
%
Intuitively, if an explanation style satisfies our desideratum of appropriate fairness perceptions, then the associated fairness perceptions should be significantly higher for the fair ADS than for the unfair ADS.
No significant differences (or reversely, higher perceptions for the unfair ADS) suggest that a given explanation style may not (sufficiently) enable people to detect the problematic behavior of the unfair ADS.
To examine whether explanations actually \emph{increase} (\emph{decrease}) fairness perceptions in case of the fair (unfair) ADS, we will also include a control group of study participants that is not provided any explanations but the introductory text in Fig.~\ref{fig:introductory_text}.
We will examine these relationships and test associated hypotheses in our main study.

\section{Conclusion and Outlook}\label{sec:outlook}

It is often argued that one goal of explaining automated decision systems (ADS) is to facilitate positive perceptions of users towards ADS.
We warn, however, that this goal should be conditional on the quality (e.g., fairness) of the system, which is---given ever more complex and opaque underlying ML models---often difficult to assess ex-ante.
To that end, we introduce a novel desideratum for explanations, \emph{appropriate fairness perceptions}, which means that an explanation should enable decision-subjects to properly assess the fairness of ADS, as indicated through their positive or negative fairness perceptions.
%
%
We then propose a randomized between-subject study design to evaluate this criterion for commonly-used explanation styles.
A key challenge in our study involves the construction of fair and unfair ADS, due to different prevalent notions of (un)fairness.
In fact, it can be assumed that different ways of constructing unfair ADS will lead to different explanation styles enabling appropriate fairness perceptions.
Ours, as well as external follow-up work, should therefore employ different notions of (un)fairness---and contexts---and compare the results.
In an immediate next step, we will conduct an online pilot study to collect preliminary data, validate our study design, and estimate certain parameters, such as the required sample size.
After that, we will initiate our main study and thoroughly discuss the findings.
We hope that our contribution will shed some new light on the effectiveness of explanations for ADS.

\begin{acks}
    The authors would like to thank the anonymous reviewers for their constructive and thoughtful feedback, as well as Simone Stumpf, Jonathan Dodge, and Min Kyung Lee, among others, for their valuable input during the \emph{Workshop on Transparency and Explanations in Smart Systems (TExSS)} at IUI '21.
\end{acks}

\bibliographystyle{ACM-Reference-Format}
\bibliography{bibliography}

\end{document}